\documentclass[prd]{revtex4}
\usepackage{epsfig}
\usepackage{mathrsfs}
\usepackage{amsmath}
\usepackage{amssymb}
\usepackage{color}
\usepackage{graphicx}

\begin{document}

\title{Charged pion chain decay and the cosmic ray positron flux}

\author{Paolo Lipari}

\email{paolo.lipari@roma1.infn.it}
\affiliation{INFN, Sezione Roma ``Sapienza'',
 piazzale A.Moro 2, 00185 Roma, Italy}

\affiliation{
 IHEP, Key Laboratory of Particle Astrophysics,
 Chinese Academy of Sciences, Beijing, China}

\affiliation{
 Tata Institute of Fundamental Research, Homi Bhabha Road, Mumbai 400005, India}

\begin{abstract}
The study of the cosmic ray positron flux has attracted intense attention
in recent years, especially because the observations suggest that it 
could receive contributions from sources such as 
pulsars or the self--annihilation or decay of dark matter particles.
The main known source of relativistic positrons, that form the background to
possible additional contributions, is the chain decay of $\pi^+$ produced
in the inelastic interactions of cosmic rays with interstellar
and circumstellar gas.
The shape of the energy spectrum of positrons produced in these pion
decays can be calculated exactly and is well known. However, surprisingly,
some estimates of the contribution of the standard mechanism
to the positron flux have adopted an incorrect spectral shape of the positron
produced in these decays, following an error present in a 1997 paper
of Moskalenko and Strong.
In this work we report this error, discuss its origin,
and estimate its impact on the numerical results. 
\end{abstract}

\maketitle

\section{Introduction}
\label{sec:intro}
Interpreting the cosmic ray (CR) positron flux is a problem
of central importance for our understanding of high--energy processes
in the Milky Way. The main known source of relativistic positrons in the
Galaxy is the secondary production mechanism, in which positrons are produced
in the decay of secondaries, mostly $\pi^+$, created in collisions
of cosmic rays with gas in interstellar space or
inside and in the vicinity of astrophysical accelerators.
However, most interpretations of the observations
\cite{PAMELA:2008gwm,AMS:2014xys,AMS:2019rhg,Fermi-LAT:2011baq}
argue that the secondary production
mechanism cannot reproduce the shape of the measured positron spectrum
because it predicts a spectral shape that is too soft.
This conclusion, often referred to  as the  existence of a ``positron anomaly'',
implies that our Galaxy contain one or more additional sources
or relativistic positrons. Possible sources  include
the self--annihilation or decay of dark matter particles \cite{Bertone:2004pz}
or the creation of electron--positron  pairs,  and their subsequent
acceleration in sources such as pulsars \cite{Hooper:2008kg}.

An alternative possibility is that the failure of the ``standard''
mechanism to reproduce the positron observations reflects the inclusion
of incorrect assumptions in constructing the predictions.
The observations could be the explained revising these assumptions.
For example assuming a shorter cosmic ray residence time
in the Milky Way and avoiding the need for an additional positron source
\cite{Cowsik:2013woa,Lipari:2016vqk}.
The implications of both types of interpretations are broad and profound.

The calculation of the positron flux due to the secondary production mechanism
includes several elements: the energy and spatial distributions
of the populations of CR protons and nuclei in the Milky Way,
the spatial distribution of target gas in interstellar space,
the inclusive spectra of secondary mesons produced in
inelastic hadronic collisions, and the properties
of propagation of relativistic charged particles
in the Galactic magnetic fields.
The possible contributions of CR collisions inside or in the envelopes of their
sources is potentially very important, but is also very uncertain,
and its inclusion  requires the inclusion of several, not well constrained,
additional parameters.

An additional element in the calculation is the description of the 
energy spectra of the positrons generated in the decay
of the particles produced in the inelastic hadronic collisions.
The dominant channel is the charged pion decay 
$\pi^+ \to \mu^+ \; \nu_\mu$ followed by the muon decay
$\mu^+\to e^+ \; \nu_e \; \overline{\nu}_\mu$.
The charged conjugate channel ($\pi^- \to \mu^- \to e^-$)
is a source of relativistic electrons, 
however  it is subdominant  compared  to the contribution
of electrons accelerated in sources.

The shapes of the  spectra of particles generated in the chain decay of
charged pions  can be calculated exactly,  and are well known
\cite{Gaisser:2016uoy,Lipari:1993hd}, 
however several papers and numerical estimates  of  the CR positron flux
have adopted  an incorrect shape for the positron spectrum,
following an error contained  in an article published
by Moskalenko and Strong in 1997
\cite{Moskalenko:1997gh} (MS97 in the following).
This paper is one of the first detailed calculation of the
$e^\pm$ spectra generated by the secondary production
mechanism, and has been cited many times
(and continues to be cited) in the literature that discusses
the CR positron flux.
In the MS97 paper appendix~C gives explicit expressions
for the spectra of the $e^\pm$ generated in $\pi^\pm$ decay,
that differ from each other. This is incorrect, because the
$e^\pm$ spectra from $\pi^\pm$ decay are in fact identical.
A difference in shape between the $e^+$ and $e^-$ spectra
would imply a large violation of $CP$ invariance,
(with $C$ and $P$ the charge conjugation and parity operators),
in conflict with the well established fact that
CP violation effects are small, confined to only few phenomena,
and absent in the $\pi^\pm$ decays.
The $e^-$ spectrum described in the MS97 paper is correct,
but the $e^+$ one is too hard.

The goal of this paper is to note this error, 
discuss its origin, and evaluate its effects on the numerical results.
The effects are only small, and in fact negligible compared to the other 
uncertainties in the calculation, nonetheless it is desirable to correct this error,
that is conceptually serious, and has been a source of confusion.

In the next section we outline a derivation of the spectra of the final state particles
in charged pion chain decay. In section~\ref{sec:power-law}  we discuss the spectra
of the $e^\pm$  and neutrinos generated by the decay of a population of charged pions 
with an energy distribution of power--law form. The final section gives a brief
summary and conclusions.

\section{Energy spectra in pion chain decay}
\label{sec:spectra}
Charged pions decay with with a branching ratio close to 100\%
into the two body final state 
$\pi^+ \to \mu^+ \, \nu_\mu$ ($\pi^- \to \mu^- \, \overline{\nu}_\mu$).
In the parent rest frame the two final state particles
are monochromatic,  with  equal and back to back.
momenta and energies:
\begin{equation}
 E_\mu^* = \frac{m_\pi}{2} \, (1+r) ~~, ~~ E_\nu^* = \frac{m_\pi}{2} \, (1-r)
\end{equation}
with: $r = (m_\mu/m_\pi)^2 \simeq 0.57309$.

Pions are scalar particles and in the pion rest frame
the total angular momentum of the final state must vanish.
This requires (in the parent rest frame)
that $\mu^\pm$ must have helicity $h = \mp 1$ to cancel the angular momentum
of the accompanying $\nu_\mu(\overline{\nu}_\mu)$,
that also have helicity $\mp 1$.
This is the well known reason behind the suppression of the
$e^+\nu_e$ ($e^-\overline{\nu}_e$) decay mode. A charged lepton
$\ell^\pm$ created by the Weak interaction has chirality
$\chi = \pm 1$, and therefore a probability
of order $(m_\ell /E)^2$ to be in the ``wrong''  (opposite sign) helicity state.
This results in a suppression of order $(m_e/m_\mu)^2$ of the $e$--like mode
with respect to the $\mu$--like mode.
The polarization of the muons created in $\pi^\pm$ decay is important
in determing the energy spectra of the particles produced in the
subsequent three body decays of the muons
($\mu^+ \to e^+ \, \nu_e \overline{\nu}_\mu$ and charge conjugate mode).

Muon decay is a purely leptonic process mediated by the Weak interactions,
and its properties can be exactly calculated.
In the rest frame of the muon,
the inclusive distributions of the three types of particles produced in the
decay of a $\mu^\pm$ with spin pointing in a fixed direction
have the form:
\begin{equation}
 \left . \frac{dN}{dx \, d\cos\theta^*}
 \right |_{\mu^\pm \to a (\overline{a})} 
= F_0^a (x) \pm \cos \theta^* \, F_{\rm pol}^a (x) 
\label{eq:mudecx}
\end{equation}
where $x = 2 E^*/m_\mu$ is an adimensional quantity proportional to the
c.m. energy $E^*$ that takes values in the interval [0,1],
and $\theta^*$ is the emission angle with respect to the spin of the parent muon.
These spectra manifestly violate parity $P$ and charge conjugation $C$,
but are also manifestly symmetric for a $CP$ transformation.

In leading order, and neglecting the electron mass,
the spectra (normalized to unit area) have the explicit forms:
\begin{equation}
 \left . \frac{dN}{dx \, d\cos \theta^*} 
 \right |_{\mu^\pm \to e^\pm} =
 \left . \frac{dN}{dx \, d\cos \theta^*} 
 \right |_{\mu^\pm \to \overline{\nu}_\mu (\nu_\mu)} 
 = (3 - 2 x) \, x^2 \mp \cos \theta^* \,(1- 2 x) \, x^2
~~~,
 \label{eq:mudec1}
\end{equation}
\begin{equation}
 \left . \frac{dN}{dx \, d\cos\theta^*} 
 \right |_{\mu^\pm \to \nu_e (\overline{\nu}_e)} =
6 \, (1 - x) \,x^2 \;
(1 \mp \cos \theta^*)
\label{eq:mudec2}
\end{equation}
In this approximation ($m_e \to 0$) 
the spectra of $e^\pm$ and $\nu_\mu(\overline{\nu}_\mu)$ are identical.

Integrating in $x$, the average values of the emission angle (with respect to
the muon spin) are:
\begin{equation}
 \langle \cos \theta^* \rangle_{e^\pm} =
 \langle \cos \theta^* \rangle_{\overline{\nu}_\mu (\nu_\mu)}
 = \pm \frac{1}{9} ~~~,
\end{equation}
\begin{equation}
 \langle \cos \theta^* \rangle_{\nu_e (\overline{\nu}_e)}
 = \mp \frac{1}{3}~~ ~.
\end{equation}
This indicates that in the decay of a positive muon ($\mu^+$)
the $e^+$ and $\overline{\nu}_\mu$
are preferentially emitted in the direction of the muon spin:
$\langle \cos \theta^* \rangle = +1/9$, 
while the $\nu_e$ are emitted preferentially in the opposite
direction: $\langle \cos \theta^* \rangle = -1/3$.
These results can be understood qualitatively as a simple 
consequence of angular momentum conservation.
In a Weak decay anti--particles (such as $e^+$ and $\overline{\nu}_\mu$)
are emitted with chirality $\chi = +1$
and therefore (for small $m/E$) also in a state of
helicity close to $+1$.
Conservation of angular momentum therefore favors their emission
in directions where their spin tends to be parallel to the
 spin of the parent particle.
The opposite is true for the $\nu_e$ that are created with
helicity $h = -1$ and preferentiall emitted in the direction opposite to
the muon spin.

In the decay of negative muons the situation is reversed,
$e^-$ and $\nu_\mu$ are preferentially emitted in the direction opposite to
the parent spin, while $\overline{\nu}_e$ preferentially toward
the spin direction.

\subsection{Pion rest frame spectra}
\label{sec:rest-frame}
The calculation of the inclusive spectra of the final state particles
after chain decay is now a straightforward exercise that simply requires
to perform Lorentz transformations, taking into account for the muon
polarization.

The calculation is very simple for the pion rest frame,
where the $\mu^\pm$ are isotropic, monochromatic, and have a unique
value of the helicity.

It is convenient to describe the spectra in the pion rest frame
as a function of the adimensional variable:
$y = 2 E/m_\pi$, that takes values in the interval $[0,1]$,
and can be expressed in terms of variables
in the muon rest frame
($x = 2 E^*/m_\mu$ and $\cos\theta^*$) as:
\begin{equation}
y = \frac{x}{2} \; \left [ (1+ r) \mp (1-r) \, \cos \theta^* \right ]~.
\label{eq:ypi}
\end{equation}
In this equation the $\mp$ sign is for $\pi^\pm$ decay,
and we have used the fact that the angle $\theta^*_\beta$ (with
respect to the muon velocity) is related to the angle
$\theta^*$ (with respect
to the pion spin) by: $\cos \theta^*_\beta = \mp \cos\theta^*$,
(because the $\mu^\pm$ helicity is $h = \mp 1$),
and the muon velocity is $\beta_\mu = (1-r)/(1+r)$.

The pion spectra can then be obtained with the integration:
\begin{eqnarray}
 \frac{dN}{dy} & = & 
 \int_0^1 dx ~\int_{-1}^{+1} ~d\cos\theta^{*}
 ~\frac{dN}{dx \, d\cos\theta^*} 
 ~\delta \left [ y - \frac{x}{2} \, [(1+r) \mp (1-r) \, \cos \theta^{*} ]
 \right ]
 \nonumber \\[0.15 cm]
 & = & \frac{2}{1-r} \, \int_y^{{\rm min}[1,y/r]} 
 \left . \frac{dx}{x} ~ \frac{dN}{dx \, d\cos\theta^*} 
 \right |_{\cos \theta^{*} =
 \mp \left (\frac{2}{1-r} \, \frac{y}{x} - \frac{1+r}{1-r}\right)}
\label{eq:gcalc}
\end{eqnarray}
The important point is to note that the inclusive spectra
for the decays
$\pi^+ \to \mu^+ \to a$ and
$\pi^- \to \mu^- \to \overline{a}$ are identical because the opposite
sign in the expression for the muon decay spectrum
[see in Eq.~(\ref{eq:mudecx}) the term $\pm \cos \theta^* \, F_{\rm pol} (x)$]
cancels the opposite sign due to the different
helicities of the $\mu^\pm$ in the pion rest frame.

\begin{figure}
\begin{center}
\includegraphics[width=14cm]{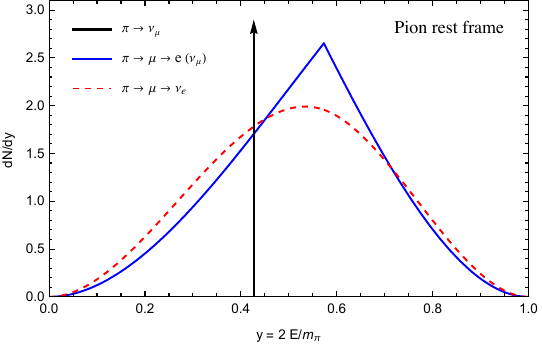}
\end{center}
\caption {\footnotesize Energy spectra (in the rest frame of the parent)
 of the particles created in the chain decay of charged pions:
 $\pi^\pm \to \mu^\pm ~\nu_\mu (\overline{\nu}_\mu)$
 followed by $\mu^+ \to e^+ \, \nu_e \, \overline{\nu}_\mu$
 (or charge conjugate mode). 
 The spectra are calculated neglecting the
 electron mass, and shown as a function of
 the adimensional variable $y = 2 \, E/m_\pi$. 
 The different lines are
 for $e^ \pm$, and $\nu_e (\overline{\nu}_e)$. 
 The spectra of the $\overline{\nu}_\mu (\nu_\mu)$ created in
 the $\mu^\pm$ decay are identical to the $e^\pm$ spectra.
 The vertical line shows the (monochromatic)
 spectrum of the $\nu_\mu (\overline{\nu}_\mu)$
 created in the first pion decay.
\label{fig:spect_pirest}
} 
\end{figure}

Explicit expressions for the inclusive spectra
in the pion rest frame (normalized to unity) are: 
\begin{eqnarray}
& ~ & \left . \frac{dN}{dy}
 \right |_{\pi^\pm\to \nu_\mu (\overline{\nu}_\mu)}
 = \delta [y - (1-r)]
\label{eq:chain0}
 \\[0.35cm] & ~ &
\left . \frac{dN}{dy} \right
 |_{\pi^\pm\to \mu^\pm \to e^\pm} = 
 \left . \frac{dN}{dy} \right
 |_{\pi^\pm\to \mu^\pm \to \overline{\nu}_\mu (\nu_\mu)} = 
 \begin{cases}
 {2\, (1+2 r)\, (3 -2 y) \, y^2} / ({3 \, r^2})
 &~~{\rm for} ~~0\le y < r \\[0.15 cm]
 {2\,(3 -2 r) \,(1 - y)^2 \,(1 + 2 y)}/ [{3 \, (1-r)^2}]
 &~~{\rm for} ~~r \le y \le 1
 \end{cases}
\label{eq:chain1}
 \\[0.35cm] & ~ &
 \left . \frac{dN}{dy} \right
 |_{\pi^\pm\to \mu^\pm \to \nu_e (\overline{\nu}_e)} = 
\begin{cases}
 {4\, y^2  \, [r \, (3- 2 y) - y]} / {r^2} 
 &~~{\rm for} ~~0\le y < r \\[0.15 cm]
 {4\,(1 -y)^2 \,[y (3 - 2 r) - r]} /{(1-r)^2}
 &~~{\rm for} ~~r \le y \le 1
 \end{cases}
\label{eq:chain2}
\end{eqnarray}
These spectra are shown in Fig.~\ref{fig:spect_pirest}.
Average values of the (fractional) energy for the different
type of particles are (for $\mu^+$ decay):
\begin{eqnarray}
 \{
 \langle y \rangle_{\nu_\mu} 
 ~,~ \langle y \rangle_{e^+} 
 ~, ~\langle y \rangle_{\overline{\nu}_\mu}
 ~, ~\langle y \rangle_{\nu_e} \}
 & = &
 \{ 1-r ~ , ~ \frac{3 + 4 r}{10} ~, ~ \frac{3+ 4 r}{10} ~, ~ \frac{2+r}{5} \}
 \nonumber \\[0.23cm]
& \simeq & 
 2 \; \{0.213455
 ~,~ 0.26462
 ~,~ 0.26462
 ~,~ 0.257309\} ~.
\label{eq:zmed}
\end{eqnarray}
Energy conservations ensures that $\sum_j \langle y \rangle_j = 2$. 
One can note after muon decay the four particles generated in pion chain decay
have approximately (but not exactly) the same average energy of order $m_\pi/4$.

Taking into account for the muon polarization
results in softer spectra for $e^\pm$ and $\nu_\mu(\overline{\nu}_\mu)$
and harder spectra for $\nu_e(\overline{\nu}_e)$  compared 
to a calculation that neglects the polarization.
This is because in the first (second) case the particles
are preferentially emitted in the opposite (same)
direction of the momentum of the $\mu^\pm$.
This happens for both $\pi^+$ and $\pi^-$ decay, because the
angular distributions of the final state particles
after the $\mu^\pm$ decay are related by a
mirror reflexion (on a plane orthogonal to the muon spin),
but the $\mu^\pm$ have opposite helicities.

\subsection{Spectra in frames where the pion is ultrarelativistic}
\label{sec:ultra}
It is straightforward to compute the energy spectra of the final state
particles in an arbitrary frame where 
the parent pion has energy $E_\pi$ (and velocity $\beta_\pi$).
The spectra, expressed as a function of the
fractional energy $z = E/E_\pi$, can be calculated 
as a convolution of the energy and (isotropic) angular distribution of the final state particles
in the pion rest frame:
\begin{eqnarray}
 \frac{dN}{dz} (z, \beta_\pi) & = &
 \int_0^1 dy ~\int_{-1}^{+1} ~\frac{d\cos\theta}{2} ~\frac{dN}{dy}
 ~\delta \left \{z - \frac{y}{2} \, [1 + \beta_\pi \, \cos \theta ]
 \right \}
 \nonumber \\[0.15 cm]
 & = & \frac{1}{\beta_\pi} ~
 \int_{(2 z)/(1+ \beta_\pi)}^{{\rm Min}[1,(2z)/(1-\beta_\pi)]} \frac{dy}{y} ~
 ~\frac{dN}{dy} 
\label{eq:hcalc}
\end{eqnarray}

\begin{figure}
\begin{center}
\includegraphics[width=14cm]{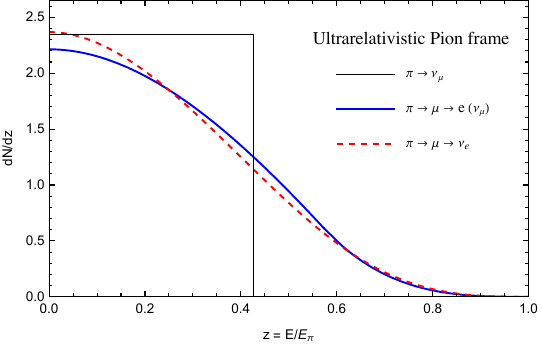}
\end{center}
\caption {\footnotesize Spectra of the final state particles
 in the chain decay of charged pions.
 The spectra are calculated in a frame where the pion is ultrarelativistic
 ($\beta_\pi \to 1$) and plotted as a function of the ratio $E/E_\pi$.
 \label{fig:spect_pilab}}
\end{figure}

In the limit of ultrarelativistic pions ($\beta_\pi \to 1$) the inclusive
spectra $dN/dz$ are independent of energy.
\begin{equation}
 \left . \frac{dN}{dz} \right |^{\beta_\pi \to 1}
 = 
 \int_z^{1} 
\frac{dy}{y} ~ \frac{dN}{dy } 
\label{eq:hultra}
\end{equation}
Explicit expressions for these asymptotic inclusive spectra are:
\begin{eqnarray}
& & \left . \frac{dN}{dz} \right
 |_{\pi^\pm\to \nu_\mu (\overline{\nu}_\mu)}^{\beta_\pi \to 1} = 
 \begin{cases}
 (1-r)^{-1}
 &~~{\rm for} ~~0\le z < (1-r) \\[0.15 cm]
 0 
 &~~{\rm for} ~~(1-r) \le z \le 1
 \end{cases}
\\[0.4cm]
& & \left . \frac{dN}{dz} \right
 |_{\pi^\pm\to \mu^\pm \to e^\pm}^{\beta_\pi \to 1}
 = 
 \left . \frac{dN}{dz} \right
 |_{\pi^\pm\to \mu^\pm \to \overline{\nu}_\mu (\nu_\mu)} = 
 \begin{cases}
 - \frac{(1+2 r) \, z^2}{r^2}
 + \frac{(4+8r) \, z^3}{9 \,r^2}
 - \frac{2(3- 2 r) \, \ln r}{3(1-r)^2}
 -\frac{2} {3 (1-r)}
 &~~{\rm for} ~~0\le z < r \\[0.15 cm]
 (2 r -3) [(4 z -9) z^2 + 6 \, \ln z + 5]/[9 (1-r)^2 \, r^2] 
 &~~{\rm for} ~~r \le z \le 1
 \end{cases}
\label{eq:z1}
\\[0.4cm]
 & & \left . \frac{dN}{dz} \right
 |_{\pi^\pm\to \mu^\pm \to \nu_e (\overline{\nu}_e)}^{\beta_\pi \to 1}
 = 
 \begin{cases}
 \frac{2 z^2 [ r (4 z -9) + 2 z]}{3 r^2}
 + \frac{4}{1-r}
 + \frac{4 r \ln r}{(1-r)^2}
 &~~{\rm for} ~~~ 0\le z < r \\[0.15 cm]
2 \, \{ r \, [( 4 z -9) z^2+ 5 ] + 6 r \ln z + 6 (1-z)^3\}/ [3 \, (1-r)^2]
 &~~{\rm for} ~~~ r \le z \le 1
 \end{cases}
\label{eq:z2}
\end{eqnarray} 
The inclusive spectra for ultrarelativistic pion frames
are shown in Fig.~\ref{fig:spect_pilab}.
The average energy values for these distributions are: 
$\langle z \rangle = \langle y \rangle/2$ [see Eq.~(\ref{eq:zmed})],
and $\sum_j \langle z \rangle_j = 1$.

The expressions of the spectra for an arbitrary value of the
pion velocity are straightforward to obtain, however because of they
take different functional forms in different $z$ intervals
because of the form of the integration limits
in Eqs.~(\ref{eq:gcalc}) and~(\ref{eq:hcalc}).
The results are shown in the appendix.
Some examples of the distributions of the decay $\pi^\pm \to \mu^\pm \to e^\pm$
for different values of $\beta_\pi$ are shown in Fig.~\ref{fig:spect_beta}.

\subsection{Spectra for wrong polarization effects}
\begin{figure}
\begin{center}
\includegraphics[width=12.9cm]{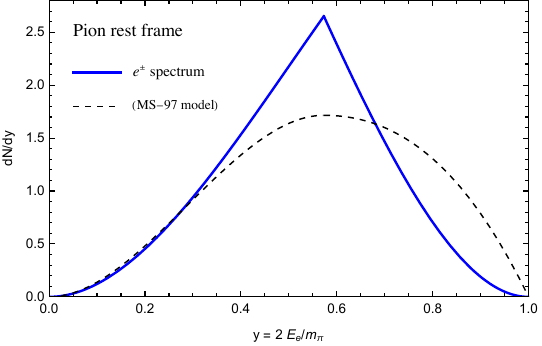}

\vspace{0.50cm}
\includegraphics[width=12.9cm]{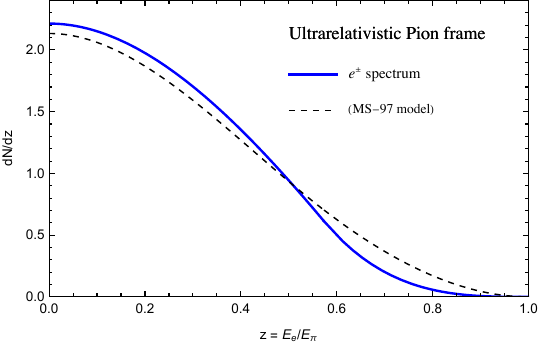}
\end{center}
\caption {\footnotesize The top panel shows as a solid line
 the spectrum of the $e^\pm$
 (and also the $\nu_\mu$ or $\overline{\nu}_\mu$)
 emitted in the chain decay of charged pions
 ($\pi^\pm \to \mu^\pm \to x$)
 plotted as a function of the adimensional variable
 $y = 2 \, E/m_\pi$. The dashed line is the positron spectrum
 used in reference \cite{Moskalenko:1997gh}.
 The bottom panel shows the same spectra for a frame where
 the parent pion is ultrarelativistic (plotted as a function
 of the quantity $z = E/E_\pi$.
 \label{fig:wrong_spectra}}
\end{figure}

The pion chain decay spectra $\pi^\pm \to \mu^\pm \to e^\pm$
described in the MS97 paper are calculated including the matrix element
for muon decay, and the dependence of the spectra on the emission angle
with respect to the muon spin, but neglect the fact 
that the $\mu^\pm$ are produced (in the pion rest frame)
with opposite polarization. This results in different spectra for the
electrons and positrons produced in $\pi^\pm$ chain decay.
This is manifestly a large violation of CP symmetry.
The electron spectrum is calculated correctly, but the positron spectrum is
a little too hard. The MS97 spectra are shown together with the
correct calculation in Fig.\ref{fig:wrong_spectra}.
Conceptually the error is serious, but the numerical effects are small. The average value of 
fractional energy of the $e^\pm$ produced in  pion  chain decay  is 
$ \langle E_{e^\pm}/E_\pi \rangle  \langle z \rangle =(3 + 4 r)/20 \simeq 0.264618$.
The MS97 positron spectrum
gives for positrons the incorrect value  $ \langle z \rangle =(4 + 3 r)/20 \simeq 0.285963$
that is 8\% higher.

\section{Positrons and neutrinos from a power--law pion spectrum}
\label{sec:power-law}
The observable fluxes of $e^\pm$ and neutrinos produced  by the the secondary
production mechanism are generated  by parent particles that in general have
very broad  energy distributions with, in first approximation, a power--law form.
This follows from the fact that the interacting cosmic rays have energy distributions of power--law form,
and the inclusive spectra of  the secondary particles produced in  their inelastic collisions
satisfy (in reasonably good approximation)  Feynman scaling \cite{Gaisser:2016uoy}.
The resulting spectra  of $e^\pm$ and neutrinos have then  power--law form with the same
slope of the parent pions (also equal to the slope of the interacting primaries).

In fact, it is well known, and straightforward to demonstrate, that 
a spectrum of ultrarelativistic charged pions of power law form
with spectral index $p$
($N_{\pi^\pm} (E_\pi) = K_{\pi^\pm} \, E_\pi^{-p}$)  that all decay, 
results in  $e^\pm$ and neutrinos  spectra   that are 
power--laws of the  the same spectral index:
\begin{equation}
N_j (E) = K_{\pi^\pm} \, Z_{\pi^\pm \to j} (p) ~E^{-p} ~.
\end{equation}
The adimensional factors $Z_{\pi^\pm \to j} (p)$ are
the $(p-1)$ momenta of the
asymptotic  distributions $dN/dz$ (in  an ultrarelativistic frame):
\begin{equation}
 Z_{\pi^\pm \to j} (p) = \int_0^1 dz \; z^{p-1}
 ~ \left [ \frac{dN}{dz} \right ]_{\pi^\pm \to j}^{\beta_\pi \to 1} 
 ~.
\label{eq:zdef}
\end{equation}
Inserting the expressions for the inclusive decay spectra  given  above one can obtain
simple  explicit expressions for the $Z$--factors:
\begin{eqnarray}
& ~ & Z_{\pi^\pm \to \nu_\mu (\overline{\nu}_\mu)} (p) = 
\frac{(1-r)^{p-1}}{p}
\\[0.3cm]
& ~& Z_{\pi^\pm \to \mu^\pm \to e^\pm} (p) = 
Z_{\pi^\pm \to \mu^\pm \to \overline{\nu}_\mu (\nu_\mu)} (p) = 
\frac{4\,\{[p (r -1) + 2 r -3] \, r^p -2 r + 3\} }{p^2\, (p+2) \, (p+3)\, (1-r)^2}
\\[0.3cm]
& ~& 
Z_{\pi^\pm \to \mu^\pm \to \nu_e (\overline{\nu}_e)} (p) = 
\frac{24\, [ r (r^p -1) + p \, (1-r) ]}{p^2\, (p+1) \, (p+2)\, (p+3) \, (1-r)^2}
\end{eqnarray} 

From their definition in Eq.~(\ref{eq:zdef}) it is easy to see  that:
For $p=1$ the $Z$ factor gives the average multiplicity of particle $j$ in the final state,
and therefore $Z_{\pi \to j} (1)= 1$ for all particle types;
and for $p=2$ the $Z$ factors are equal to the average
$\langle z \rangle_j = \langle E_j \rangle /E_\pi$, that is the fraction of the
parent pion energy carried on average by particles of type $j$
in the final state (see Eq.~(\ref{eq:zmed}), as it is straightforward to  verify.

\begin{figure}
\begin{center}
\includegraphics[width=14cm]{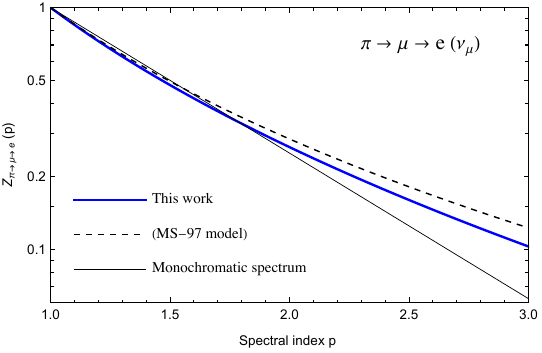}
\end{center}
\caption {\footnotesize The thick solid line shows
the $Z$--factor that corresponds to the spectra of
the $e^\pm$ and $\nu_\mu$ ($\overline{\nu}_\mu$) produced
in the chain decay $\pi^\pm \to \mu^\pm \to x$.
The dashed line is the $Z$-factor obtained using the incorrect 
spectrum for the positrons emitted the chain pion decay
in reference \cite{Moskalenko:1997gh}.
The thin line shows the $Z$--factor that corresponds to
describing the spectrum as a sharp line 
at energy $E = E_\pi/4$.
 \label{fig:spect_momenta}}
\end{figure}

Fig.~\ref{fig:spect_momenta} shows the
the $Z$--factor for the  decay $\pi^\pm \to \mu^\pm \to e^\pm$
as a function of the spectral index $p$ [see Eq.~(\ref{eq:z1})], 
 and compares with the result obtained for
the incorrect positron spectrum of the MS97 paper.
The results are also compared with what is obtained 
adopting the often used approximation of a monochromatic decay spectrum:
$dN/dz = \delta[z-1/4]$.
All $Z$--factors are unity,  and  are therefore independent from the
shape of the decay spectra  for $p = 1$.
The differences between the calculations become  more imnportant  for soft pion  spectra
(larger  spectral indices), but remain always modest.

\begin{figure}
\begin{center}
\includegraphics[width=14cm]{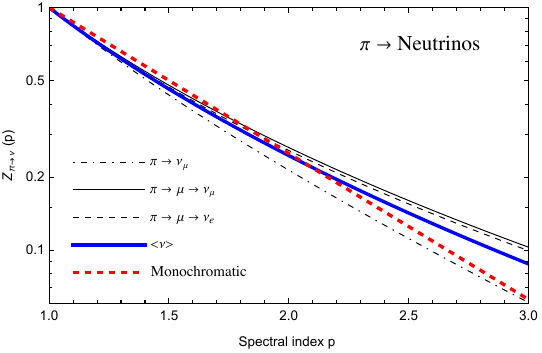}
\end{center}
\caption {\footnotesize The thin lines show 
the $Z$--factors that corresponds to the spectra of
the three neutrinos emitted in the chain decays of charghed pions.
The thick solid line is the average of the three curves.
The thick dashed line corresponds to the approximation
of describing the spectra as monochromatic
with energy $E_\nu = E_\pi/4$.
 \label{fig:nu_momenta}}
\end{figure}

Figure~\ref{fig:nu_momenta} shows the
$Z$--factors for the decay of charged pions into neutrinos of different type,
showing also the average (relevant because of the effects of flavor oscillations).
The results are again also compared with the  approximation of monchromatic  decay spectra
($E_\nu = E_\pi/4$ for all neutrinos). Also in this case the differences are more
important for large values of the spectral index, but remain modest.

A precise description of the decay spectra is clearly relevant also
to interpret  neutrino spectra that deviate  from a simple power--law form.

\section{Summary and conclusions}
In this work we have collected analytic expressions
for the spectra of the final state particles emitted in
the chain decay of charged pions (an $e^\pm$ and three neutrinos).
These  spectra have been obtained in leading order and neglecting the electron mass,
and also assuming that the muons created in the initial pion decay
do not lose energy before their decay, and that their helicity does not change
during their propagation. These approximations are in general very good for
applications of astrophysical interest.

One motivation for this paper is to correct an error, present
in an often cited paper of Moskalenko and Strong \cite{Moskalenko:1997gh},
that states that the spectra for the decays
$\pi^+ \to \mu^+\to e^+$
and
$\pi^- \to \mu^- \to e^-$
are different.
This error has certainly been noticed before by many readers,
but no correction has yet appeared in the literature,
and the error had not been recognized by the authors.
Since  the MS97 paper  remains an important and often cited reference
for the  calculation for the positron, correcting the error eliminates
a  source of confusion.
This error is conceptually serious, but its numerical effects for  the predictions of the 
cosmic ray positron spectrum are only modest, and correspond to
an overestimate of order 10\%, that is very small when compared
to other uncertainties, therefore our discussion is essentially
only of conceptual interest.

In this paper, for completeness, we have also included
expressions for the spectra of the neutrinos produced in charged pion decay,
and have discussed the order of magnitude of the errors introduced
by the commonly used approximations of
describing the spectra of $e^\pm$ and neutrinos 
as monochromatic (with energy $E = E_\pi/4$), for the very simple,
but phenomenologically interestig case of a power--law
spectrum of the decaying pions.
The error depends on the spectral index of the decaying pions
but remains below the 10--15\% level.

\vspace{0.5 cm}

\noindent{\bf Acknowledgments}\\[0.15 cm]
Discussions  with Andrew Strong and Igor Moskalenko are gratefully acknowledged.

\vspace{0.5cm}
\appendix

\section{Energy spectra in the laboratory frame}
In this appendix we give expressions for the inclusive energy
spectra of the final state particles in $\pi^\pm$ chain decay
in an arbitrary frame where the pion has velocity $\beta_\pi$.
The spectra are expressed in terms of the fractional energy
$z = E/E_\pi$.

For the neutrinos emitted in the 2--body initial pion decay one has
the ``box--like'' form:
\begin{equation}
 \left [\frac{dN}{dz} (z, \beta_\pi)
 \right ]_{\pi^\pm \to \nu_\mu (\overline{\nu}_\mu)} = 
 \begin{cases}
 0 & ~{\rm for:}~~~ z < (1-r)
 \left ( \frac{1-\beta_\pi}{2} \right ) \\[0.3cm]
 [(1-r)\beta_\pi]^{-1} & ~{\rm for:}~~~ (1-r)
 \left( \frac{1-\beta_\pi}{2} \right ) \le z \le
 (1-r)\left (\frac{1+\beta_\pi}{2} \right ) \\[0.3cm]
 0 & ~{\rm for:}~~~ z > (1-r) \left ( \frac{1+\beta_\pi}{2} \right ) 
\end{cases}
\end{equation}

The spectra of the particles emitted in the subsequent muon decay
have different functional forms in different sub-intervals of $z$, and
to obtain more compact expressions 
it is convenient to introduce the functions:
\begin{equation}
 A_{l,h} (y) = \int \frac{dy}{y} ~\left [\frac{dN}{dy} \right ]_{y < r \; (y > r)}
\end{equation}
where the function $dN/dy$ is the
spectrum of the particle under consideration in the pion rest frame
(with $y = 2 E^*/m_\pi$, and $E^*$ the rest frame energy).
The $dN/dy$ functions (given in the main text in section~\ref{sec:ultra})
have different functional forms in the low ($y < r$)
and high ($y > r$) $y$ range, and the subscript indicates if
the integral is calculated for the low or high $y$ form.
Explicit expressions for these functions are:
\begin{eqnarray}
 & A_l^{\pi^\pm \to \mu^\pm \to e^\pm} (y) =
 A_l^{\pi^\pm \to \mu^\pm \to \overline{\nu}_\mu (\nu_\mu)} (y) = &
\frac{(1+2 r) \, y^2(9 - 4 y)}{9  r^2}
 + {\rm const.}
 \\[0.15cm]
 & A_h^{\pi^\pm \to \mu^\pm \to e^\pm} (y) =
 A_h^{\pi^\pm \to \mu^\pm \to \overline{\nu}_\mu (\nu_\mu)} (y) = &
\frac{(3-2 r) \, [(4 y -9) y^2  + 6 \ln y]}{9 (1-r)^2} 
 + {\rm const.}
\end{eqnarray}
and
\begin{eqnarray}
 & A_l^{\pi^\pm \to \mu^\pm \to \nu_e (\overline{\nu})_e} (y) = &
2 y^2[r (9- 4y) - 2 y]/(3 r^2) 
  + {\rm const.}
\\[0.15 cm]
& A_h^{\pi^\pm \to \mu^\pm \to \nu_e (\overline{\nu})_e} (y) = &
\frac{12 y - 6 (2 -r)y^2 + (4/3) (3-2 r) y^3 - 4 r \ln y}{(1-r)^2}
 + {\rm const.}
\end{eqnarray}
The $dN/dz$ spectra can be then expressed as combinations of
the $A_{l(h)}$ functions. One has to consider two ranges of
the pion velocities: 
($\beta_\pi \le \beta_\mu^*$) and 
($\beta_\pi > \beta_\mu^*$), where 
\begin{equation}
\beta_\mu^* = \frac{(1-r)}{(1+r)} \simeq 0.271383
\end{equation}
is the velocity of the muons in the pion rest frame.

For $\beta_\pi \le \beta_\mu^*$ one has:
\begin{equation}
\frac{dN}{dz} (z, \beta_\pi) = 
 \begin{cases}
 \frac{1}{\beta_\pi} \left \{
 A_l\left (\frac{2 z}{1- \beta_\pi} \right )
 -A_l\left (\frac{2 z}{1+\beta_\pi} \right ) \right \}
 & ~{\rm for:}~~~ 0\le z < \frac{1}{2} \, (1-\beta_\pi) \, r
 \\[0.15 cm]
 \frac{1}{\beta_\pi} \left \{
 A_l\left (r \right )
 -A_l\left (\frac{2 z}{1+\beta_\pi} \right )
 +A_h\left (\frac{2 z}{1-\beta_\pi} \right )
 -A_h\left (r \right )
 \right \}
 & ~{\rm for:}~~~
 \frac{1}{2} \, (1-\beta_\pi) \, r \le z <
 \frac{1}{2} \, (1+\beta_\pi) \, r 
 \\[0.15 cm]
 \frac{1}{\beta_\pi} \left \{
 A_h\left (\frac{2 z}{1-\beta_\pi} \right ) 
 -A_h\left (\frac{2 z}{1+\beta_\pi} \right ) 
 \right \}
 & ~{\rm for:}~~~
 \frac{1}{2} \, (1+\beta_\pi) \, r \le z <
 \frac{1}{2} \, (1- \beta_\pi)
 \\[0.15 cm]
 \frac{1}{\beta_\pi} \left \{
 A_h\left (1 \right ) 
 -A_h\left (\frac{2 z}{1+\beta_\pi} \right ) \right \}
 & ~{\rm for:}~~~
 \frac{1}{2} \, (1-\beta_\pi) \le z <
 \frac{1}{2} \, (1+\beta_\pi) 
 \\[0.15 cm]
 0 & ~{\rm for:}~~~
 z > \frac{1}{2} \, (1+\beta_\pi) 
 \end{cases}
\end{equation}
For $\beta_\pi > \beta^*_\mu$ one has:
\begin{equation}
\frac{dN}{dz} (z, \beta_\pi) = 
 \begin{cases}
 \frac{1}{\beta_\pi} \left \{
 A_l\left (\frac{2 z}{1- \beta_\pi} \right )
 -A_l\left (\frac{2 z}{1+\beta_\pi} \right ) \right \}
 & ~{\rm for:}~~~ 0\le z < \frac{1}{2} \, (1-\beta_\pi) \, r
 \\[0.15 cm]
 \frac{1}{\beta_\pi} \left \{
 A_l\left (r \right )
 -A_l\left (\frac{2 z}{1+\beta_\pi} \right )
 +A_h\left (\frac{2 z}{1-\beta_\pi} \right )
 -A_h\left (r \right )
 \right \}
 & ~{\rm for:}~~~
 \frac{1}{2} \, (1-\beta_\pi) \, r \le z <
 \frac{1}{2} \, (1-\beta_\pi) 
 \\[0.15 cm]
 \frac{1}{\beta_\pi} \left \{
 A_l\left (r \right )
 -A_l\left (\frac{2 z}{1+\beta_\pi} \right ) 
 +A_h\left ( 1 \right ) 
 -A_h\left (r \right ) \right \}
 & ~{\rm for:}~~~
 \frac{1}{2} \, (1-\beta_\pi) \le z <
 \frac{1}{2} \, (1+ \beta_\pi)\, r 
 \\[0.15 cm]
 \frac{1}{\beta_\pi} \left \{
 A_h\left (1 \right ) 
 -A_h\left (\frac{2 z}{1+\beta_\pi} \right ) \right \}
 & ~{\rm for:}~~~
 \frac{1}{2} \, (1+\beta_\pi) \, r \le z <
 \frac{1}{2} \, (1+\beta_\pi) 
 \\[0.15 cm]
 0 & ~{\rm for:}~~~
 z > \frac{1}{2} \, (1+\beta_\pi) 
 \end{cases}
\end{equation}
In the limit of $\beta_\pi \to 1$ the first two and the last ranges
in the previous equation shrink to zero, and one  obtains:
\begin{equation}
\frac{dN}{dz} (z, \beta_\pi\to 1) = 
 \begin{cases}
 A_l\left (r \right )
 -A_l\left ( z \right ) 
 +A_h\left ( 1 \right ) 
 -A_h\left (r \right ) 
 & ~{\rm for:}~~~
 0 \le z < r 
 \\[0.15 cm]
 A_h\left (1 \right ) 
 -A_h\left (z \right ) 
 & ~{\rm for:}~~~
z > r
 \end{cases}
\end{equation}

\begin{figure}
\begin{center}
\includegraphics[width=14cm]{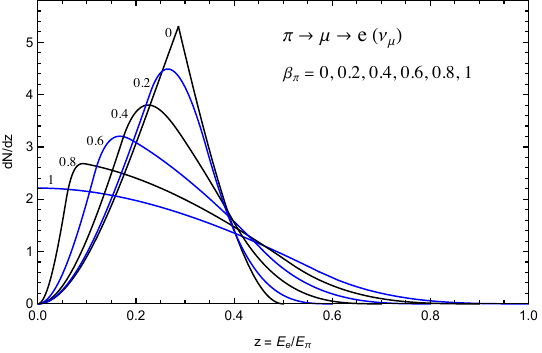}
\end{center}
\caption {\footnotesize Energy spectra of the electrons and positrons
 created in the chain decay $\pi^\pm \to \mu^\pm \to e^\pm$.
 The spectra are calculated in a ``laboratory frame'' where the parent pion
 has velocity $\beta_\pi = 0$, 0.2, 0.4, 0.6, 0.8 and 1, and plotted as
 a function of the ratio $z = E_e/E_\pi$.
 \label{fig:spect_beta}}
\end{figure}

Figure~\ref{fig:spect_beta} shows the $e^\pm$ spectra
in frames where the parent pion has velocity
$\beta_\pi = 0$, 0.2, 0.4, 0.6, 0.8 and 1.


\clearpage

\begin{thebibliography}{100}

\bibitem{PAMELA:2008gwm}
O.~Adriani \textit{et al.} [PAMELA],
``An anomalous positron abundance in cosmic rays with energies 1.5-100 GeV,''
Nature \textbf{458}, 607-609 (2009)
doi:10.1038/nature07942
[arXiv:0810.4995 [astro-ph]].


\bibitem{AMS:2014xys}
M.~Aguilar \textit{et al.} [AMS],
``Electron and Positron Fluxes in Primary Cosmic Rays Measured with the Alpha Magnetic Spectrometer on the International Space Station,''
Phys. Rev. Lett. \textbf{113}, 121102 (2014)
doi:10.1103/PhysRevLett.113.121102

\bibitem{AMS:2019rhg}
M.~Aguilar \textit{et al.} [AMS],
``Towards Understanding the Origin of Cosmic-Ray Positrons,''
Phys. Rev. Lett. \textbf{122}, no.4, 041102 (2019)
doi:10.1103/PhysRevLett.122.041102


\bibitem{Fermi-LAT:2011baq}
M.~Ackermann \textit{et al.} [Fermi-LAT],
``Measurement of separate cosmic-ray electron and positron spectra with the Fermi Large Area Telescope,''
Phys. Rev. Lett. \textbf{108}, 011103 (2012)
doi:10.1103/PhysRevLett.108.011103
[arXiv:1109.0521 [astro-ph.HE]].

\bibitem{Bertone:2004pz}
G.~Bertone, D.~Hooper and J.~Silk,
``Particle dark matter: Evidence, candidates and constraints,''
Phys. Rept. \textbf{405}, 279-390 (2005)
doi:10.1016/j.physrep.2004.08.031
[arXiv:hep-ph/0404175 [hep-ph]].

\bibitem{Hooper:2008kg}
D.~Hooper, P.~Blasi and P.~D.~Serpico,
JCAP \textbf{01}, 025 (2009)
doi:10.1088/1475-7516/2009/01/025
[arXiv:0810.1527 [astro-ph]].



\bibitem{Cowsik:2013woa}
R.~Cowsik, B.~Burch and T.~Madziwa-Nussinov,
``The origin of the spectral intensities of cosmic-ray positrons,''
Astrophys. J. \textbf{786}, 124 (2014)
doi:10.1088/0004-637X/786/2/124
[arXiv:1305.1242 [astro-ph.HE]].


\bibitem{Lipari:2016vqk}
P.~Lipari,
``Interpretation of the cosmic ray positron and antiproton fluxes,''
Phys. Rev. D \textbf{95}, no.6, 063009 (2017)
doi:10.1103/PhysRevD.95.063009
[arXiv:1608.02018 [astro-ph.HE]].

 

\bibitem{Gaisser:2016uoy}
T.~K.~Gaisser, R.~Engel and E.~Resconi,
``Cosmic Rays and Particle Physics''
Cambridge University Press, 2016,
ISBN 978-0-521-01646-9


\bibitem{Lipari:1993hd}
P.~Lipari,
``Lepton spectra in the earth's atmosphere,''
Astropart. Phys. \textbf{1}, 195 (1993)
doi:10.1016/0927-6505(93)90022-6

\bibitem{Moskalenko:1997gh}
I.~V.~Moskalenko and A.~W.~Strong,
(MS97), 
Astrophys. J. \textbf{493}, 694-707 (1998)
doi:10.1086/305152
[arXiv:astro-ph/9710124 [astro-ph]].

\end{thebibliography}
\end{document}